\documentclass[twocolumn,showpacs,preprintnumbers,amssymb,pra]{revtex4}
\usepackage{graphicx}
\usepackage{slashbox}
\usepackage{dcolumn}
\usepackage{bm}
\usepackage{latexsym,epsfig}

\begin{document}
\title{Reliability test for the experimental results of electric quadrupole hyperfine structure constants and new assessment of nuclear quadrupole moments in $^{135,137}$Ba}
\author{B. K. Sahoo$^1$ \footnote{Email: bijaya@prl.res.in}, M. D. Barrett$^2$ and B. P. Das$^3$}
\affiliation{$^1$Theoretical Physics Division, Physical Research Laboratory, Ahmedabad-380009, India}
\affiliation{$^2$Centre for Quantum Technologies and Department of Physics, National University of Singapore, 3 Science Drive 2, 117543 Singapore}
\affiliation{$^3$Theoretical Astrophysics Group, Indian Institute of Astrophysics, Bangalore-560034, India}
\date{Received date; Accepted date}

\begin{abstract}
We investigate electric quadrupole hyperfine structure constant ($B$) results
in $^{135}$Ba$^+$ and $^{137}$Ba$^+$ and compare their ratios with the reported
measurements of the ratio between the nuclear quadrupole moment ($Q$) values of these isotopes.
We carry out confidence tests for the reported experimental $B$ values and
calculations of $B/Q$ from the present work.
Inconsistencies in the experimental $B$ values are observed in both the
isotopes from different experiments performed using the same techniques. The present
calculations are carried out using an all order relativistic many-body
theory considering only single and double excitations in the
coupled-cluster ansatz. After a detailed analysis of the results, the values of
$Q$ we obtain for $^{135}$Ba and $^{137}$Ba are 0.153(2)$b$ and 0.236(3)$b$,
respectively, which differ by about 4\% from the currently referred precise values.
\end{abstract}

\maketitle

The effective ellipsoidal nuclear charge distribution of an isotope 
in the nuclear ground state 
is generally quantified by the nuclear electric quadrupole moment ($Q$).
The nuclear shell model assumes that the charge distribution is spherically
symmetric for a doubly-magic closed-shell nucleus, and therefore its
quadrupole moment vanishes; otherwise 
it will have an intrinsic non-zero $Q$ value.
Thus, accurate evaluations of $Q$ for different isotopes could
probe the shell theory (e.g. see \cite{haas}) and therefore the investigation
of this quantity has been of considerable interest to nuclear, atomic and molecular
physicists for the last six decades \cite{ramsey,feld,dufek,tanaka,minamisono,neyens}.

Although the value of $Q$ for nuclear ground states cannot be measured directly, it is possible to measure their ratios for different isotopes \cite{williams,sternheimer,voss,keim}. Also, for an atomic state of a particular isotope, $Q$ can be determined by
combining the measured electric quadrupole hyperfine structure constant ($B$)
with a calculation of $B/Q$ for that isotope. To obtain an accurate value of $Q$ by this procedure, both the measurement and the calculation have to be performed to high precision. Experimental results are generally considered to be more reliable than the calculated results for many-electron atoms owing to the fact that {\it ab initio} evaluation of various physical quantities using many-body
methods involve a number of approximations at different stages of the calculations. However, measurements from different experiments are not always in agreement and, in certain cases, large discrepancies between measurements and accurate calculations have been noted \cite{sahoo1,sahoo2}. It is therefore essential to scrutinize the accuracies of both the experimental $B$ and calculated $B/Q$ results when determining values of $Q$ from them.

\begin{table*}[t]
\caption{Calculated results of $R^e=B/Q$ in MHz/$b$ and $\Delta$ factors in $^{137}$Ba$^+$ and $^{135}$Ba$^+$ from the present work. Estimated uncertainties from the calculations are given in the parentheses.}\label{tab1}
\begin{center}
\begin{ruledtabular}
\begin{tabular}{l|l|ccc|cc}
Atomic &  \multicolumn{4}{c|}{$^{137}$Ba} & \multicolumn{2}{c}{$^{135}$Ba} \\
\cline{2-5} \cline{6-7} State & \backslashbox{$R^e_{137}$}{$\Delta_{137}^{137}$} & $[5p^6] \ 6p_{3/2}$ & $[5p^6] \ 5d_{3/2}$ & $[5p^6] \ 5d_{5/2}$ & $\Delta_{135}^{137}$  & $R^e_{135}$ \\
\hline \\
$[5p^6] \ 6p_{3/2}$ & 379.52(4.0) & 0.0 & $-0.50(1)$ & $-0.33(2)$ & $-1.2\times10^{-7}$ & 379.52(4.0)  \\
$[5p^6] \ 5d_{3/2}$ & 188.69(2.5) & 1.01(5) & 0.0 & 0.35(3) & $3.0\times 10^{-7}$ & 188.69(2.5) \\
$[5p^6] \ 5d_{5/2}$ & 253.76(3.5) & 0.50(4) & $-0.26(2)$ &  0.0 & $-4.0\times 10^{-7}$ & 253.76(3.5) \\
\end{tabular}
\end{ruledtabular}
\begin{tabular}{lc}
\end{tabular}
\end{center}
\end{table*}
Studies of hyperfine structure constants in singly ionized barium (Ba$^+$) can provide valuable information about the
wave functions in the nuclear region and the role of the electron correlation effects whose knowledge are relevant for the proposed parity nonconservation experiments \cite{sahoo5}. It is also important for the extraction of higher order nuclear moments \cite{lewty}. Furthermore, studies of the hyperfine structure are also interesting for the investigation of the variation of the fine structure constant \cite{karshenboim} and for astrophysical studies \cite{mashonkina}.

In this Rapid Communication, we investigate the consistency between measured values of $B$ and compare these to calculations of $B/Q$ for the 135 and 137 isotopes of Ba$^+$. This provides us with new estimates of the quadrupole moments for these two isotopes that differ from previous estimates by approximately $4\%$.  In addition we show that there are significant discrepancies in the experimentally measured values of $B$.

The nuclear electric quadrupole hyperfine structure constant $B$ for an atomic state with angular momentum $J > 1/2$ is defined as \cite{schwartz}
\begin{eqnarray}
B &=& Q \left \{ \frac {8J(2J-1)}{(2J+1)(2J+2)(2J+3)} \right \}^{1/2} \langle J|| \textbf{T}^{(2)}||J\rangle \nonumber \\
 &=& Q R^e(J)
\label{eqn1}
\end{eqnarray}
The operator $\textbf{T}^{(2)}=\sum t^{(2)}$ is a second rank tensor whose reduced
matrix element in terms of the single particle orbitals is given by
\begin{eqnarray}
\langle \kappa_f || t^{(2)} || \kappa_i \rangle &=& \frac{(-1)^{j_f-1/2}}{2} \left [1- \frac{\kappa_f \kappa_i}{|\kappa_f \kappa_i|} (-1)^{j_f+j_i} \right] \nonumber \\ && \sqrt{(2j_f+1)(2j_i+1)} \left ( \matrix { j_f & 2 & j_i \cr 1/2 & 0 & -1/2 \cr } \right ) \nonumber \\ &&  \int_0^{\infty} dr \frac{(P_fP_i+Q_fQ_i)}{r^3}, \ \ \ \
\label{eqn2}
\end{eqnarray}
where $\kappa_i$ and $j_i$ are the angular momentum quantum numbers of $i^{th}$
Dirac orbital with large $P_i$ and small $Q_i$ radial components. In Eq. (\ref{eqn1}),
we have defined the ratio $R^e(J)=B/Q$ which is a function of $J$ in the electronic 
coordinate and can be determined directly from theoretical calculations.

Within first order perturbation theory, the hyperfine structure constants for a particular state are linearly related to measurements of the hyperfine splittings (see, for example, \cite{arimondo} and references therein).  Thus, provided higher order terms can be neglected, the hyperfine structure constant, $B$, can be determined directly from experiments. Measured values of $B$ and calculated values of $R^e$ can then be used to determine $Q$.  Since the ratio of $Q$ values between isotopes can be measured directly, a useful consistency check on both measurements and calculations is to consider ratios between different isotopes or different states within the same isotope.  In general we have the relationship
\begin{eqnarray}
\left ( \frac{B_1}{B_2} \right )^\mathrm{expt}&=& \left ( \frac{Q_1}{Q_2}\right )^\mathrm{expt} [1 + \Delta^1_2]^\mathrm{theor},
\label{eqn3}
\end{eqnarray}
where we have introduced $\Delta^1_2=(R^e_1-R^e_2)/R^e_2$ and the subscripts 
1 and 2 denote the relevant parameter for different states and/or isotopes.  
The superscripts $\mathrm{expt}$ and $\mathrm{theor}$ are used to emphasize 
the fact that the corresponding term can be obtained from experimental 
measurements and theoretical calculations, respectively.  The parameter 
$\Delta^1_2$ is analogous to the hyperfine anomaly that has been studied 
extensively for the magnetic dipole hyperfine structure constant ($A$) 
in a number of atomic systems \cite{buttgenbach, persson}, but has not 
been studied for $B$.  Indeed, the above expression can be generalized 
to all hyperfine structure constants.

In Table~\ref{tab1}, we present calculated values of $R^e$ and $\Delta$ among the $6p_{3/2}$,
$5d_{3/2}$ and $5d_{5/2}$ states for $^{135}$Ba$^+$ and $^{137}$Ba$^+$. Atomic states that have been considered here have a common closed core $[5p^6]$ and one electron in the valence orbital, denoted by $k$, in the different quantum states in Ba$^+$.
The wave functions of these states with the Dirac-Coulomb (DC) Hamiltonian are
calculated using the relativistic coupled-cluster (RCC) method by expressing them 
in a form \cite{geetha}
\begin{eqnarray}
| \Psi_k \rangle &=& e^T \{1+S_k\} | \Phi_k \rangle ,
\label{eqn5}
\end{eqnarray}
where $|\Phi_k \rangle$ is the Dirac-Fock (DF) reference state corresponding to the valence
electron $k$. $T$ and $S_k$ are the operators that include correlations among the core electrons
and the valence and the core electrons respectively. We consider all possible single and double
excitations (known as CCSD method) along with the leading triple excitations (known as CCSD(T) method) involving the valence orbital.

The finite charge distribution of the nucleus in the different isotopes is modeled by taking a
two-parameter Fermi-nuclear charge distribution
\begin{equation}
\rho(r_i) = \frac {\rho_0} {1 + e^{(r_i-c)/a}},
\label{eqn6}
\end{equation}
where $\rho_0$ is the density for the point nuclei, $c$ and $a$ are the half-charge radius and skin
thickness of the nucleus. These parameters are chosen as $a = 2.3/4(ln3)$ and
$c = \sqrt{ \frac{5}{3} r_{rms}^2 - \frac{7}{3} a^2 \pi^2}$, where $r_{rms}$ is the root
mean square radius of the atomic nucleus that is given in the Fermi model as
$r_{rms}= 0.836{\cal A}^{1/3} + 0.57$ in fermi ($fm$) with atomic mass $\cal{A}$. To reduce the
uncertainty in our calculation for different charge distributions,
we consider the optimal values of $r_{rms}$ as 4.8273 and 4.8326 $fm$
for $^{135}$Ba$^+$ and $^{137}$Ba$^+$, respectively, as tabulated in \cite{angeli}.

\begin{table}[h]
\caption{Reported ratios of $Q$ values from various experiments. $Q$ values are given in $b$.}\label{tab2}
\begin{ruledtabular}
\begin{tabular}{ll}
$Q$($^{137}$Ba)/$Q$($^{135}$Ba) & Method \\
\hline \\
 $0.245(4)/0.160(3)=1.531(38)$ & Fast beam spectroscopy of \\
                               & $6p_{3/2}$ state in Ba$^+$ \cite{wendt} \\
 $0.228(24)/0.146(16)=1.56(24)$ & Level-crossing spectroscopy \\
                                & of $5d 6p \ ^3P_1$ state in Ba  \cite{ma} \\
 $0.246(2)/0.150(15)=1.64(16)$ & Fast beam spectroscopy of \\
                               & $5d_{3/2}$ state in Ba$^+$ \cite{silverans}\\
 $0.248(3)/0.162(2)=1.531(26)$ & Non-relativistic calculation \\
                                  & using CCSD method \cite{martensson} \\
                         1.543(3) & Proton magnetic resonance \\
                                  & resonance in BaBr$_2$ \cite{williams}\\
                         1.537(2) & Level-crossing spectroscopy \\
                                  & of $5d 6p \ ^3P_1$ state of Ba \cite{putlitz} \\
                         1.49(10) & NMR spectroscopy of \\
                                  & BaCl$_2$ \cite{kruger} \\
                         1.538485(95) & NMR and NQR spectroscopy \\
                                      & of BaCl$_2$ \cite{lutz} \\
                         1.5426(62)   & Microwave spectroscopy \\
                                      & of BaO \cite{blom} \\
\end{tabular}
\end{ruledtabular}
\end{table}

\begin{table}[t]
\caption{Brief summary of the measurements of $B$ (in MHz) in different states of $^{137}$Ba and $^{135}$Ba and in their singly charged ions from the measurements carried out by same experiments. Ratios of these quantities are also given to compare them with the ratio of $Q$ values listed in Table \ref{tab2}.}\label{tab3}
\begin{ruledtabular}
\begin{tabular}{lccc}
  State &  $^{137}$Ba & $^{135}$Ba & Ratio \\
\hline \\
Ba I \\
$[5p^6] 6s 5d \ ^3D_2$ & 26.8(30)$^a$    & 18.3(22)$^a$   & 1.46(24)  \\
                       & 25.899(13)$^b$  & 16.745(14)$^b$ & 1.5467(15) \\
$[5p^6] 6s 5d \ ^3D_3$ & 36(9)$^a$       & 20(8)$^a$      & 1.8(8)  \\
                       & 47.390(16)$^b$  & 30.801(24)$^b$ & 1.5386(13) \\
$[5p^6] 6s 5d \ ^1D_2$ & 59.564(14)$^a$ & 38.710(15)$^a$ & 1.5387(68) \\
$[5p^6] 6s 5d \ ^3D_1$ & 17.890(3)$^b$ & 11.642(4)$^b$ & 1.53668(59) \\
$[5p^6] 6s 6p \ ^1P_1$ & 50.09(21)$^c$    & 34.01(22)$^c$ & 1.473(11)  \\
                       & 49.7(5)$^d$      & 32.5(4)$^d$   & 1.529(24) \\

$[5p^6] 6s 6p \ ^3P_1$ & -41.61(2)$^e$   & -27.08(2)$^e$ & 1.5366(14)  \\
$[5p^6] 6p 5d \ ^3D_1$ & $-3.5(6)^f$ & $-2.4(10)^f$ & 1.46(66) \\
$[5p^6] 6p 5d \ ^3F_2$ & 45.3(10)$^f$ & 29.0(10)$^f$ & 1.562(64) \\
$[5p^6] 6p 5d \ ^3F_3$ & 83.0(14)$^f$ & 53.9(14)$^f$ & 1.540(48) \\
$[5p^6] 6p 5d \ ^3F_4$ & 111.0(13)$^f$ & 71.1(13)$^f$ & 1.561(34) \\
$[5p^6] 6s 7p \ ^1P_1$ & 16.6(6)$^g$ & 8.8(6)$^g$ & 1.89(15) \\
 & & \\
Ba II \\
$[5p^6] 6p_{3/2}$ & 92.5(2)$^h$ & 59.0(1)$^h$ & 1.5678(43) \\
                  & 95.0(37)$^i$ & 59.4(23)$^i$ & 1.599(88) \\
                  & 89.7(15)$^j$ & 55.3(34)$^j$ & 1.62(10) \\
                  & 91.6(12)$^k$ & 62.1(12)$^k$ & 1.475(34) \\
                  & 92.5(4)$^l$ & 60.4(4)$^l$ & 1.531(12) \\
                  & 97.1(93)$^l$ & 63(9)$^l$ & 1.54(27) \\
$[5p^6] 5d_{3/2}$ & 47.5(13)$^j$    & 33.2(24)$^j$    &  1.43(11) \\
                  & 44.5408(17)$^m$ & 28.9528(25)$^m$ & 1.53839(15) \\
$[5p^6] 5d_{5/2}$ & 80.7(10)$^j$    & 45.2(10)$^j$    & 1.785(45) \\
                  & 59.533(43)$^m$  & 38.692(10)$^m$  & 1.5386(12) \\
\end{tabular}
\end{ruledtabular}
\begin{tabular}{lcccccccc}
References: & $^a$\cite{schmelling}; & $^b$\cite{gustavsson}; & $^c$\cite{wijngaarden}; &
            $^d$\cite{nowicki}; & $^e$\cite{putlitz}; & $^f$\cite{grundevik}; & $^f$\cite{eliel}; \\
 & $^h$\cite{villemoes}; & $^i$\cite{wendt1}; & $^j$\cite{silverans1}; & $^k$\cite{winter}; & $^l$\cite{andrei}; & $^m$\cite{silverans} \\
\end{tabular}
\end{table}

The uncertainties in the calculations presented in Table~\ref{tab1} are estimated from the finite size of bases, approximations in the level of calculations and higher order relativistic corrections.  As seen in the table, the $\Delta_{137}^{137}$ factors with respect to different
states are large and the associated uncertainties are also significant. However, the
$\Delta_{135}^{137}$ factor for a state is below $10^{-6}$ and it can only play a role when
the accuracies of the quantities of interest are comparable to these values. This is due to the fact that the wave function for a particular state does not depend significantly on the isotope and we can expect this to be true for the neutral atom also. Thus, from Eq. (\ref{eqn3}), we see that the ratio $B_{137}/B_{135}$ gives the ratio $Q_{137}/Q_{135}$ which is independent of the state of interest.  This provides a direct consistency check of the experimentally measured ratios of $B$ and $Q$.

\begin{figure}[t]
  \includegraphics[scale=0.6]{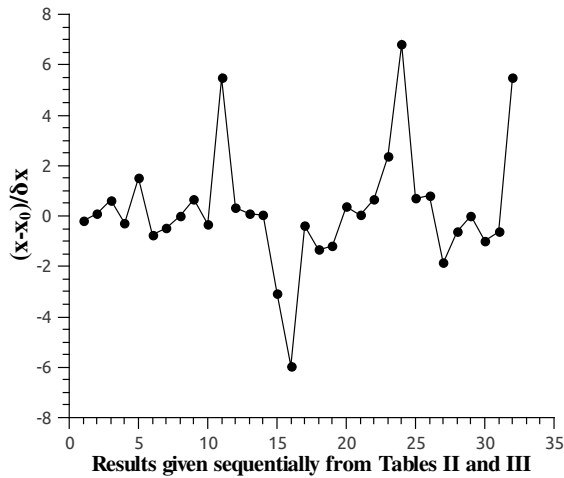}
 \caption{Plot showing $(x-x_0)/ \delta x $ with $x$ representing ratio of $Q$ values for 
$^{137}$Ba and $^{135}$Ba obtained from different works given sequentially in Tables \ref{tab2}
and \ref{tab3} (taken in X-axis), $x_0$ denoting the most accurately reported ratio 
$1.538485$ in Ref. \cite{lutz} and $\delta x$ correspond to errors associated with the 
respective $x$ values.}
  \label{fig1}
\end{figure}

In Table~\ref{tab2} we present reported measurements of the ratio $Q_{137}/Q_{135}$. The last 
four are measured directly using nuclear magnetic resonance techniques whereas the first four 
are extracted from hyperfine measurements. In Table \ref{tab3}, we present 
the reported measurements of $B$ in both atoms and singly charged ions of $^{135}$Ba and 
$^{137}$Ba, along with the ratio $B_{137}/B_{135}$. The uncertainties given are assumed to 
represent one standard deviation. To highlight the variation in the results, in 
Fig.~\ref{fig1}, we plot the deviation of each ratio relative to its respective uncertainty 
from the most accurately reported value of 1.538485. From this plot we see there are four 
experimental values in which the deviation is more than 5 standard deviations. Thus these 
reported values should be treated as questionable. We note that a $\chi^2$ minimization of the 
remaining ratios gives a reduced $\chi^2$ of 1.02 and an estimated ratio of 1.538425(78) 
consistent with the reported value of 1.538485(95). We therefore conclude that the reported 
value of 1.538485(95) is a reliable estimate of the ratio $Q_{137}/Q_{135}$.

Having established a reliable value of $Q_{137}/Q_{135}$, we now proceed to assess the validity of the calculations presented in Table~\ref{tab1}.  To do this we consider experimentally measured $B$ values for different states within the same isotope, specifically $^{137}$Ba$^+$.  From Eq. (\ref{eqn3}), it is seen that the ratio of these $B$ values give a direct measure of the calculated value of $\Delta_{137}^{137}$.  Thus comparison of the measured ratio of $B$ values with $1+\Delta_{137}^{137}$ provides a consistency check for our calculated $R^e$ values.  Taking the most accurately reported values of $B$ in Table~\ref{tab3}, namely those for the $d_{3/2}$ and the $d_{5/2}$ states from \cite{silverans}, we deduce a value of $1+\Delta_{137}^{137}=1.3366(1)$ consistent with our calculated value of $1.35(3)$.  Thus we may now reliably use the measured values of $B$ with our calculated values of $R^e$ to deduce $Q$ values for the two isotopes.

To obtain the greatest accuracy, we determine the $Q$ value using data from the $d_{3/2}$ state of $^{137}$Ba$^+$. Recently the $B$ value for this state has been measured to very high accuracy $44.5387936(10)$ MHz \cite{lewty} consistent with the value given in Table~\ref{tab3}.  Combined with our calculated $R^e$ value given in Table~\ref{tab1} we obtain $Q$ for $^{137}$Ba of 0.236(3) $b$. Further, from the ratio of $Q$ values, 1.538485(95), we then deduce a value of $Q$ for $^{135}$Ba of 0.153(2) $b$. Comparing these values with other reported values of $Q$ in Table
\ref{tab2}, we find significant differences in the results. Recently, the value of $Q$ for $^{137}$Ba was evaluated by us to be 0.246(1) $b$ \cite{bijaya}. This value was determined using the experimental result for $B$ of the $[5p^6] 6p_{3/2}$ state from \cite{villemoes}.  However, as we have shown here, those measurements have significantly differ from the other measurements.

\begin{table}[t]
\caption{New $B$ results for $^{137}$Ba$^+$ and $^{135}$Ba$^+$ (in MHz) obtained combining the new $Q$ values with our calculations of $R^e$ factors .}\label{tab4}
\begin{ruledtabular}
\begin{tabular}{lcc}
  State &  $^{137}$Ba & $^{135}$Ba \\
\hline \\
$[5p^6] 6p_{3/2}$ & 89.57(1.48) & 58.07(1.0) \\
$[5p^6] 5d_{3/2}$ & 44.53(82)   & 28.87(54) \\
$[5p^6] 5d_{5/2}$ & 59.89(1.0) & 38.83(74) \\
\end{tabular}
\end{ruledtabular}
\end{table}

For completeness, we can use the newly obtained $Q$ values and our calculations of the $R^e$ factors given in Table \ref{tab1}, to determine the $B$ values of the $[5p^6] 6p_{3/2}$, $[5p^6] 5d_{3/2}$ and $[5p^6] 5d_{5/2}$ states in $^{137}$Ba$^+$ and $^{135}$Ba$^+$. The results are given in Table \ref{tab4} and can be compared with the results given in Table \ref{tab3}. As seen in Table \ref{tab4}, the present estimated values are certainly not the most accurate values due to the large uncertainties in our calculated $R^e$ factors, but they give consistent results for both the isotopes and states considered. For more precise values of $B$ and to determine more precise values of $Q$ it is necessary to carry out more accurate theoretical calculations.

In summary, we have carried out an extensive consistency check of reported
experimental values of the hyperfine structure constants over a range of different
states for the 135 and 137 isotopes of Barium. Using the relativistic
coupled-cluster method, we have carried out the corresponding calculations and investigated
ratios of the quadrupole moments between $^{137}$Ba$^+$ and $^{135}$Ba$^+$. From our analysis,
we have found some inconsistencies in the reported measurements. The reason for such
inconsistencies is not clear to us; one possibility is the neglect of higher order nuclear moments or higher order correction terms to the hyperfine interaction. We have also estimated $Q$ values for both the $^{137}$Ba$^+$ and $^{135}$Ba$^+$ isotopes from results that were shown to be consistent. We propose more experimental and theoretical studies of these quantities to ascertain our results.
Nonetheless, our proposed analysis for the consistency check of the experimental results will also be useful in the future studies of the above properties.

{\it Acknowledgement:}
The calculations reported in this work were performed using the 3TFLOP HPC cluster computational
facility at the Physical Research Laboratory, Ahmedabad.

\end{document}